\documentclass[prl,twocolumn,showpacs,epsfig,preprintnumbers,amsmath,amssymb]{revtex4}

\usepackage{ulem,color}
\usepackage{graphicx,amsthm,latexsym,epsfig,subfigure,bm,color}

\def \half{ \frac {1}{2} }
\def \bea{\begin{eqnarray}}
\def \eea{\end{eqnarray}}

\begin{document}
\title{Interior design of a two-dimensional semiclassic black hole}
\author{Dana Levanony and Amos Ori}
\affiliation{Department of Physics, Technion, Haifa 32000, Israel}
\date{\today}
 \begin{abstract}
We look into the inner structure of a two-dimensional dilatonic
evaporating black hole. We establish and employ the homogenous
approximation for the black-hole interior. The field equations
admit two types of singularities, and their local asymptotic
structure is investigated. One of these singularities is found to
develop, as a spacelike singularity, inside the black hole. We
then study the internal structure of the evaporating black hole
from the horizon to the singularity.
\end{abstract}

\maketitle

\section{Introduction}
The discovery of Hawking radiation and the black-hole (BH)
evaporation phenomenon \cite{Hawking} raised several outstanding
open questions and intriguing puzzles. One such problem which
attracted a lot of attention is the \textit{information puzzle}:
Simple thought experiments suggest that in the process of
black-hole formation and evaporation, a pure initial quantum state
will evolve into a mixed state, and consequently part of the
information encoded in the initial state will get lost.

Apparently this problem of information loss is intimately related
to another conceptual problem in black-hole physics: The formation
of a spacetime singularity inside the BH. Pictorially speaking, a
portion of the initial information propagates to the singularity
and disappears there. It is widely anticipated, however, that the
formation of spacetime singularities is a mere artifact of the
classical (and semiclassical) theory, but Quantum Gravity will
eventually resolve the black-hole singularities, and presumably
solve the information puzzle as well.

In 1992, Callan, Giddings, Harvey, and Strominger (CGHS)
\cite{CGHS} proposed a simplified framework for investigating
black-hole formation and evaporation. They introduced a
two-dimensional toy-model with gravity and matter fields coupled
to a dilaton scalar field. At the classical level this model
admits a one-parameter family of static black-hole solutions
(parameterized by their mass). When semiclassical corrections are
added, the two-dimensional black hole emits thermal radiation and
consequently evaporates. Based on the trace anomaly, CGHS provided
simple explicit expressions for the semiclassical contribution to
the energy-momentum tensor. Consequence the semiclassical dynamics
may be formulated as a closed (constrained) system of second-order
partial differential equations. It thus provides a simple
framework for exploring various aspects of black-hole evaporation,
and particularly the information puzzle \cite{related_work}.

Originally it was hoped \cite{CGHS} that the CGHS
evaporating  BH will be free of any singularities. However,
Russo, Susskind and Thorlacius \cite{Singularity_first_notice}
soon found that a singularity inevitably develops inside the CGHS
BH, at a certain value of the dilaton filed. Thus, the CGHS
formalism does not resolve the information puzzle at the
semiclassical level. One may still hope, however, that when the
model is fully quantized the singularity (and the information
puzzle) will be resolved. This approach was pursued by Ashtekar,
Taveras, and Varadarajan \cite{Ashtekar}, who formulated the
quantum-field analog of the two-dimensional CGHS model. In this
set-up, the original CGHS variables are replaced by quantum
operators. Spacetime evolution is then described by a system of
operator partial differential equations. The exact solution to
these operator equations is not known. Nevertheless, some
approximate solutions were constructed \cite{Ashtekar}, illuminating certain aspects of
the problem of black-hole evaporation, and
suggesting that the singularity will indeed be resolved in the
quantized theory.

The above discussion signifies the inevitable spacetime
singularity as a key feature of the CGHS semiclassical BH
spacetime. However, the detailed structure and properties of this
singularity have not been investigated so far to the best of our
knowledge. The main purpose of this manuscript is to present such
a detailed analysis of the structure of that singularity. The
motivation for this investigation is obvious from the discussion
above: First, since this singularity plays such a crucial role in
the information-loss puzzle, it will be useful to better
understand its properties. For example, one would like to know how
strong this singularity is, and what are the prospects for
extending semiclassical spacetime beyond it. Second, understanding
the asymptotic behavior of the various fields at the singularity
may provide a useful starting point for exploring how quantum
treatments (like that of Ref. \cite{Ashtekar} for example) may
resolve the singularity.

Although the main objective of this paper is the asymptotic
behavior near the singularity, we also analyze here the internal
structure of the evaporating semiclassical BH in the entire range
from the horizon to the singularity. We do this by constructing
approximate solutions in various domains of the BH interior, and
then matching these solutions at their respective overlap regions.
Understanding the entire BH internal structure is interesting on
its own right, but is also important for full determination of the
singularity structure: When the latter is derived by a purely
local analysis, one obtains a family of local asymptotic solutions
which depend on certain free parameters (or free functions,
depending on the context). The value of these parameters in an
actual BH solution needs to be determined by matching to initial
conditions (e.g. at the horizon). The full internal solution, from
the horizon to the singularity, is required for obtaining the
right values of these parameters.

As long as the evaporating black hole is macroscopic, the
semiclassical effects are very weak in a local dynamical sense.
This allows one to employ the \textit{adiabatic approximation}.
Namely, any local region of spacetime may be well approximated by
a certain classical CGHS BH solution (with a certain mass
parameter). The classical BH solutions are static outside the
horizon but homogeneous inside it, implying that the interior of a
macroscopic evaporating BH is (locally) approximately homogeneous.
This \textit{homogeneous approximation} is a key element in our
investigation. It greatly simplifies the analysis, as the field
equations now reduce to {ordinary} differential equations.

In Sec. II we present the CGHS action and field equations, and
re-formulate them in new variables. Sec. III is devoted to the
homogenous approximation. We explore its main properties and its
domain of validity. In Sec. IV we analyze the singularities of the
homogenous field equations, and obtain the local asymptotic
behavior of the solution near these singularities. In Sec. V we
construct an approximate solution for the fields inside the
evaporating BH, in the entire range from the horizon to the
spacelike singularity.

In Sec. VI we present simple effective Lagrangian and
Hamiltonian which yields the local asymptotic dynamics near the
spacelike singularity. The construction of such effective
Lagrangian and Hamiltonian is motivated by the idea, that perhaps
it will be possible to employ simple quantum-mechanical
considerations to resolve the singularity (and thereby to explore
the possible extension of spacetime beyond it). Although such a
quantum-mechanical treatment is beyond the scope of the present
paper, we nevertheless take here the preparatory step
(carried out entirely within the semiclassical framework) of
constructing the effective near-singularity Lagrangian and
Hamiltonian. Finally, in the last section we summarize our main results and discuss their significance.

\section{The model and field equations}

We begin by presenting the two-dimensional semiclassical CGHS
model \cite{CGHS}. Beside the dilaton $\phi$, this model contains
a large number $N>>1$ of identical scalar matter fields $f_{i}$,
and a cosmological constant $\lambda^2$. We express the metric in
double-null coordinates $u,v$ (the "conformal gauge"), namely
$ds^2=-e^{2\rho}dudv$. The action then reads \bea
\begin{array}{l}\frac{{\rm 1}}{\pi }\int{} d^2 \sigma
 [ e^{ - 2\phi } \left( { - 2\rho,_{uv}  + 4\ \phi,_u   \phi,_v  - \lambda ^2 e^{2\rho } } \right)\\
 \,\,\,\,\, \,\,\,\,\,\,\,\,\,\,\,\,\,\,\,-\frac{1}{2}\sum\limits_{i = 1}^N { f_i,_u  f_i,_v }+\frac{N}{{12}} \rho,_u \rho,_v ] \\
 \end{array} .\label{Action}
\eea The last term in the action expresses the semiclassical
effects. The Einstein equations take the form: \bea  T_{ uv }&=e^{
- 2\phi } \left( {2\phi,_{uv}- 4  \phi,_u  \phi,_v  - \lambda ^2
e^{2\rho } } \right)
 -\frac{N}{12}\rho,_{uv} = 0 \label{T_{uv}}
\eea and \bea T_{ uu }= e^{-2\phi}\left( {4\rho,_u \phi,_u  -
2\phi,_{uu} } \right) + \frac{1}{2}\sum\limits_{i = 1}^N { f_i,_u
f_i,_u }\\ \nonumber  -\frac{N}{12}[ \rho,_u
\rho,_u-\rho,_{uu}-\hat{z}_u(u)]= 0
,\,\,\,\,\,\,\,\,\,\,\,\,\,\,\,\,\,\,\,\,\,\,\,\,\,\,\label{CGHS_metric_eq}\eea

 \bea T_{ vv }= e^{-2\phi}\left( {4\rho,_v \phi,_v  -
2\phi,_{vv} } \right) + \frac{1}{2}\sum\limits_{i = 1}^N { f_i,_v
f_i,_v }\\ \nonumber  -\frac{N}{12}[ \rho,_v
\rho,_v-\rho,_{vv}-\hat{z}_v(v)]= 0
,\,\,\,\,\,\,\,\,\,\,\,\,\,\,\,\,\,\,\,\,\,\,\,\,\,\,\eea where
$\hat z_u \left( u \right)$ and $\hat z_v \left(v\right)$ encode
the information about the initial quantum state, which in turn
determines the semiclassical fluxes.
The dilaton and matter equations are: \bea -4\phi,_{uv}  + 4\phi,_u\phi,_v  + 2\rho,_{uv}  &+ \lambda ^2 e^{2\rho }  = 0, \label{CGHS_dilaton_matter_eq}\\
\partial_u\partial_v f_i = 0. \nonumber
 \eea

Throughout this paper we set $\lambda =1$. This choice is equivalent to the change of variable
$\rho'=\rho +\ln(\lambda)$, which does not affect the field equations otherwise.
We also set $f_{i}=0$, as we are dealing here with the evaporation rather than formation of the BH.

Following Ref. \cite{Amos_inside an evaporating 2D CBH}, we define new variables:
$R\equiv e^{-2\phi}$, $S\equiv 2(\rho-\phi)$. A straightforward
substitution of the new variables in
Eqs. (\ref{T_{uv}},\ref{CGHS_dilaton_matter_eq}) yields:
\bea &R,_{uv}  =  - e^S  - K\rho ,_{uv}, \label{evolution_eq_mixed}\\
&S,_{uv}  = K\rho ,_{uv} /R \nonumber ,
\eea
where $\rho=\frac{1}{2}(S-\ln R)$ is to be substituted,
and $K\equiv N/12$ expresses the magnitude of the quantum effects.
The constraint equations become: \bea R,_{ww}  - R,_w S,_w  +
\hat T_{ww}  = 0,\label{constraint_evolution_mixed}\eea
where hereafter $w$ stands for either $u$ or $v$.
The semiclassical energy fluxes $\hat{T}_{ww}$ along both null directions are given by:
\bea \hat T_{ww}  = K\left[ {\rho ,_{ww}  - \rho ^2 ,_w  + \hat
z_w \left( w \right)} \right]. \eea

It is useful to re-express the system of evolution
equations (\ref{evolution_eq_mixed}) in its standard form, in which
$R,_{uv}$ and $S,_{uv}$ are explicitly given in terms of
lower-order derivatives:\bea & R,_{uv}  =  - e^S \frac{{\left( {2R - K} \right)}}{{2\left( {R - K} \right)}} - R,_u R,_v \frac{K}{{2R\left( {R - K} \right)}}, \label{final_eq_simplified}\\
& S,_{uv}  = e^S \frac{K}{{2R\left( {R - K} \right)}} + R,_u R,_v \frac{K}{{2R^2 \left( {R - K} \right)}}.\nonumber
\eea
This form makes it obvious that the evolution equations become singular when $R=K$.
This singularity, which in the original variables takes place at $\phi=-\half \ln(\frac{N}{12})$,
was already noticed preivously \cite{Singularity_first_notice}.
Below we shall explore in some detail the homogeneous variant
of this singularity.

The semi-classical equations reduce to the classical ones by setting $K=0$: \bea & R,_{uv}  =  - e^S,  \label{Schwarzschild equations}\\
& S,_{uv}  = 0, \nonumber
\eea and the constraint equations: $$R,_{ww}-R,_w S,_w=0.$$
This set of equations
admits a one-parameter family of solutions (up to gauge transformations), which is the dilatonic
two-dimensional analog of the standard Schwarzschild solution.
Each member of this family describes a black-hole space-time, which
(like its standard four-dimensional counterpart) is static outside the BH and homogeneous inside it.
 Throughout this paper we shall simply refer to this class of solutions
as the "Schwarzschild solution" (despite a slight abuse of standard terminology).
We shall focus on the internal part of the BH. In Eddington-like double-null coordinates the
internal solution takes the form
\bea &R =  - e^{v+u} + M, \label{schwartz_sol_eddington}\\
 &S = v+u, \nonumber
 \eea
 where $M$ is the (dilatonic, 2-dimensional) Schwarzschild mass parameter.

\section{homogenous set-up}\label{sec. Homogenous_equation}
\subsection{Justification and domain of validity}
We shall now restrict our attention to the homogeneous solutions of the semiclassical
field equations inside the evaporating BH, namely, solutions which only depend on $t$,
where we define \bea t \equiv v+u , \quad \quad x \equiv v-u .   \nonumber \eea
We first need to discuss the justification for this homogeneous approximation and its domain of validity.
The classical interior solution (\ref{schwartz_sol_eddington}) is obviously
homogeneous as it depends on $v+u$ solely. In an evaporating BH, however, the semiclassical effects spoil
the exact homogeneity, as manifested by the drift in the BH mass. Yet, as long as the BH is macroscopic
($M>>K$), this drift is very slow. Indeed it undoubtedly has a dramatic effect over the long evaporation
time-scale---the BH eventually disappears (or at least becomes microscopic) after all. Yet, from a local point of view this drift
is of negligibly small rate. Thus, over a typical dynamical time/length scale (say, $\Delta v,\Delta u$
of order $\sim 10$), the change in $M$ is $\delta M<<M$ and may be neglected, which naturally
leads to the homogeneous approximation.

The local, relative magnitude of the semiclassical terms is of
order $K/R$, as can be seen in the evolution equations
(\ref{final_eq_simplified}). As long as the BH is macroscopic
($M>>K$), and as long as we are dealing with the portion $R>>K$ of
the BH space-time, we may apply the adiabatic point of view to the
evaporation process: Namely, in each region of space-time the
solution is approximately Schwarzschild, with a local effective
mass parameter $M$ which slowly drifts in $v$ and/or $u$. We shall
refer to this slowly varying function $M(v,u)$ as the {\it
effective mass} of the evaporating BH. (In fact, as long as the
interior of the BH is considered, the effective mass depends only
on $v$. \footnote{In the CGHS model the evaporation proceeds in a
constant rate $K/4$ (at the leading order); hence,
$M(v)=M_{init}-(K/4)(v-v_{init})$, where $M_{init}$ and $v_{init}$
respectively denote the initial mass and the moment $v$ of
collapse.}) The adiabatic approximation, wherever applicable,
automatically implies approximate homogeneity inside the
BH---simply because the Schwarzschild interior solution is
homogeneous.

It should be emphasized, however, that the adiabatic approximation does not apply in the small-$R$ region
near the singularity, where $R$ becomes comparable to $K$. In this region there are strong local semiclassical
effects, and the local solution is very different from Schwarzschild (as demonstrated in the next section).
We argue, however, that the homogeneous approximation is still valid in this small-$R$ domain.
Generally speaking, this follows from causality: Since the small-$R$ region is located at the causal
future of the moderate-$R$ interior region, it simply inherits the approximate homogeneity of the latter.

\begin{figure}[htb]
\begin{center}
\includegraphics[scale=0.45]{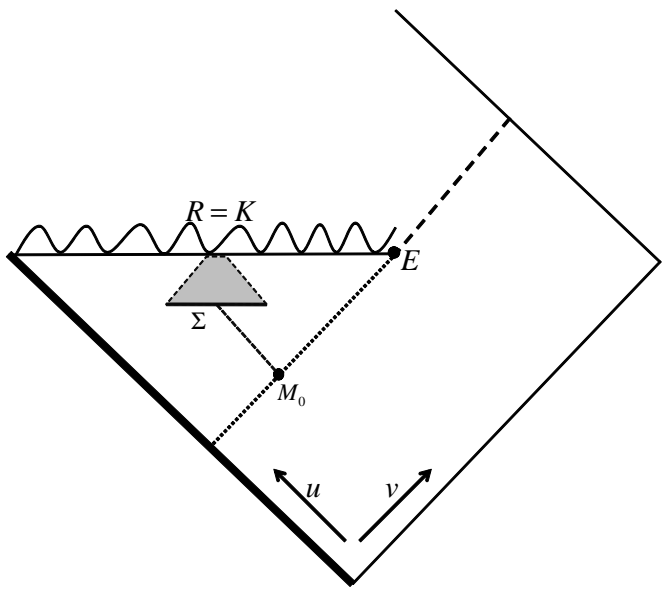}
\caption{Spacetime diagram of a CGHS black hole which forms by
gravitational collapse and subsequently evaporates. The thick null
line at the bottom left represents the collapsing massive shell.
The event horizon is marked by a dotted line. The spacelike
hypersurface $\Sigma$ is centered at $v=v_0$ (where the effective
mass is $M(v)=M_0$), and is taken to be at a certain $R$ value
satisfying $K<<R<M_0$. The gray area represents $D_{+}(\Sigma)$.
The point $E$ marks the end-point of the evaporation process. }
\label{fig_space_time}
\end{center}
\end{figure}

To be more specific, let us pick a point inside the evaporating
BH, in a region where the effective mass is still macroscopic,
$M(v)\equiv M_{0}>>K$. This point is picked at a certain $R$
value, $R=R_0$, which is $>>K$ --- say $R_{0}=M_0/2$ (recall that
the horizon is located at $R\approx M$). Let us denote the
coordinates of this point by ($v_{0},u_{0}$), and let $t_{0}\equiv
v_{0}+u_{0} , x_{0}\equiv v_{0}-u_{0}$. Let $\Sigma$ denote an
initial surface which is the line $t=t_{0}$ restricted to the
range $x_{0} \le x \le  x_{0}+\Delta x$, with $\Delta x$ taken to
be $\sim 10$ (say). This construction is illustrated in
Fig.\ref{fig_space_time}. We shall now consider two cases, namely
two slightly different solutions: In case (i), we specify on
$\Sigma$ initial data which exactly correspond to the classical
Schwarzschild solution with mass $M(v_{0})\equiv M_{0}$. Since
these data are independent of $x$ by construction, the evolving
solution will be precisely homogeneous, throughout the domain of
dependence $D_{+}(\Sigma)$. Provided that $\Delta x$ is taken to
be large enough, $D_{+}(\Sigma)$ will reach the small-$R$
space-like singularity (more precisely, the $R=K$ singularity
which we explore in the next section). To this end $\Delta x$ only
need to exceed the span of $t$ from $t_{0}$ to the singularity (of
the homogeneous solution). This span is of order $1$ (for
instance, for $R_{0}=M_{0}/2$ this $t$-span is $ln(1/2)$ in the
classical solution, and approximately the same number in the
semiclassical solution). Hence taking $\Delta x \sim 10$ (say)
guaranties that $D_{+}(\Sigma)$ will include a portion of the
small-$R$ singularity (as illustrated in
Fig.\ref{fig_space_time}).

In case (ii), we specify on $\Sigma$ the initial data which correspond to the actual evaporating BH solution.
Since both $M_{0}$ and $R_{0}$ are $>>K$, the adiabatic approximation applies,
and the solution in the neighborhood
of ($v_{0},u_{0}$) is approximately Schwarzschild. Therefore, we may regard the actual initial data (ii) as the
homogeneous, Schwarzschild data of case (i) plus a small perturbation. This initial perturbation will evolve into a small
perturbation throughout the interior of $D_{+}(\Sigma)$. The relative magnitude of this perturbation will be
controlled by the small parameter $K/M_{0}$, and for sufficiently small value of this parameter we may neglect the
perturbation at the leading order.

The smallness of the perturbations in the interior of $D_{+}(\Sigma)$ is then guaranteed by
standard stability theorems for nonlinear hyperbolic systems.
One may be concerned, however, about the effect of the perturbation in the immediate neighborhood of the small-$R$ singularity,
and in particular on the very structure of the latter.
A close examination, which is beyond the scope of the present paper, reveals that this singularity is locally
stable to inhomogeneities.
(That is, an inhomogeneous variant of the $R=K$ singularity exists \cite{Ori unpublished},
and this inhomogeneous singularity is generic, in the sense that it depends on four arbitrary functions of $x$
\footnote{The main effect of the inhomogeneity is to slowly drift the
parameters of the homogeneous solution
[e.g. the parameters $B$ and $t_{0}$ in Eq. (\ref{Solution_at_K_R})]
in the spatial direction along the singular line. The local temporal structure is not affected at the leading order.
This was verified analytically by expanding the inhomogeneous singular solution off the moment of
singularity $t_{0}(x)$ (at which $R=K$) up to seven orders in $t-t_{0}(x)$ \cite{Ori unpublished}}.)

We conclude that in the macroscopic-mass domain ($M_{0}>>K$) the homogeneous solution provides a good approximation
not only in the domain $R>>K$, but also up to the small-$R$ singularity. This was also verified
numerically by directly integrating the field equations (\ref{evolution_eq_mixed}) with initial data corresponding to a BH which forms by a
collapsing null shell and subsequently evaporates \cite{L. Dori}. In particular it was
numerically verified that the structure of the evolving spacelike singularity is well described
by the homogeneous $R=K$ singularity (described in the next section).

In the above construction we considered for concreteness an initial hypersurface $\Sigma$
located at $R=M_{0}/2$.
We could have started instead at any other $R=R_{0}$ initial value which satisfied $K<<R_{0}<M_{0}$.
The evolving homogeneous solution depends very weakly on $R_{0}$,
because of the local smallness of the semiclassical corrections at $R>>K$.
For concreteness, and in order to avoid the arbitrariness
associated with the extra parameter $R_{0}$, in the analysis below we shall consider the
homogeneous solution obtained at the limit $R_{0} \to M_{0}$.
The solution obtained in this way may be thought of as the
(horizon-regular) semiclassical counterpart of the homogeneous interior Schwarzschild solution.
\footnote{Note that the above argument which establishes the homogeneous approximation does not
directly apply to the case $R_{0} = M_{0}$, because the latter hypersurface in Schwarzschild is null rather than spacelike.
Nevertheless, the limiting solution obtained as $R_{0} \to M_{0}$ is well defined and is perfectly
regular at the horizon.}
It carries (besides $K$) a single parameter $M_{0}$, representing the BH's remaining mass
at the epoch of interest.
In the rest of the paper we shall simply refer to this parameter as $M$ for brevity.
\subsection{Homogenous field equations}
In the homogeneous framework the evolution equations (\ref{final_eq_simplified}) become \bea & \ddot{R}  =  - e^S \frac{{\left( {2R - K} \right)}}{{2\left( {R - K} \right)}} - \dot{R}^2 \frac{K}{{2R\left( {R - K} \right)}},
\label{final_homo_eq_simplified}\eea
\bea & \ddot{S}  = e^S \frac{K}{{2R\left( {R - K} \right)}} + \dot{R}^2
\frac{K}{{2R^2 \left( {R - K} \right)}},\label{final_homo_eq_simplified_S} \eea
where an over-dot denotes differentiation with respect to $t$. The constraint equation (\ref{constraint_evolution_mixed}) now reads \bea \ddot{R}  - \dot{R}\dot{S}+ \hat T  = 0, \label{semi-classical constraint}\eea
where $\hat{T}\equiv\hat{T}_{vv}=\hat{T}_{uu}$ is given by \bea \hat{T}= K [\ddot{\rho} - \dot{\rho}
^2+\hat{z}], \label{constarint}\eea
and
$\hat{z}\equiv\hat{z}_{v}= \hat{z}_{u}$.
Note that the equalities $\hat{T}_{vv}=\hat{T}_{uu}$ and $\hat{z}_{v}= \hat{z}_{u}$ are direct
consequences of the homogeneous set-up. The latter equality also implies that $\hat{z}$ must be a constant.
(By homogeneity $\hat{z}$ could at most depend on $t$,
but then $d\hat{z}/dt=\hat{z}_{,v}=\hat{z}_{u,v}=0$.)

 The Schwarzschild solution inside the BH may be expressed in the explicitly homogeneous form \bea \label{schwartzschild_sol_eq} &R =  - e^{t} + M, \\
 &S =t. \nonumber
 \eea
In particular it satisfies the relations \bea \dot{R}=-e^{S}=R-M,
\label{Schwarzschild_sol_derivative}\eea
which will be useful in the analysis below.
Note also the classical homogeneous constraint equation $\ddot{R}-\dot{R}\dot{S}=0$.

\section{Singularities}

The semiclassical equations are singular at the points $R=0$ and $R=K$,
as can be seen from Eqs. (\ref{final_homo_eq_simplified},\ref{final_homo_eq_simplified_S}).
In this section we will look at the asymptotic behavior
near these two special $R$ values.

Since we set the initial conditions for the homogeneous solution at $R>>K$, the first singularity to be encountered is $R=K$.
As it turns out, this singularity is characterized by the divergence of $\dot{R}$ and $\dot{S}$, while $R$ and $S$ are finite.

\begin{figure}[htb]
\begin{center}
\includegraphics[scale=0.6]{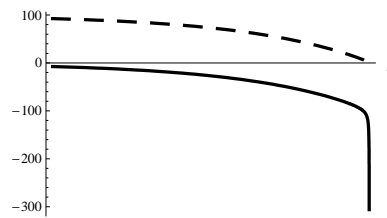}
\caption{A numerical plot of R (dashed)
and of $\dot{R}$ (solid) near the $R=K$ singularity, for $K=1$ and
$M=10^2$. While $R$ is bounded, the
divergence of $\dot R$ at the singularity is clearly seen.} \label{fig_R_Rprime}
\end{center}
\end{figure}

We therefore assume that in the right-hand side of
Eqs. (\ref{final_homo_eq_simplified},\ref{final_homo_eq_simplified_S}) we
can neglect the terms proportional to $e^S$ compared to those $\propto \dot{R}^2$, obtaining at leading order
\bea &\ddot{R}=- \frac{\dot{R}^2}{2(R - K)}, \label{IS_equations}\eea
\bea &\ddot{S}=\frac{\dot{R}^2}{2K(R - K)}. \label{Eq_IS_S} \eea
Eq. (\ref{IS_equations}) constitutes a closed equation for $R$, which is easily solved:\bea  &R\left( t \right) =K+B\left|{t - t_0 } \right|^{\frac{2}{3}}.   \label{Solution_at_K_R}\eea
 Then Eq. (\ref{Eq_IS_S}) is solved to yield \bea &S\left( t \right) =  - \frac{B}{K}\left| {t - t_0 } \right|^{\frac{2}{3}}
 + (t-t_0) B_2  + B_3.  \label{Solution_at_K_S}\eea
The solution depends on four free parameters $B,B_2,B_3,t_0$ as required, hence this is a locally-generic asymptotic solution.
We see that the variables $R,S$ are continuous at $R=K$, but $\dot{R}$
diverges as $|t - t_0|^{-1/3}$ (and the same for $\dot{S}$).
This behavior, which is numerically demonstrated in Fig. \ref{fig_R_Rprime},
justifies our preassumption that $e^S$ is indeed negligible compare to $\dot{R}^2$.

In the second singular point $R=0$, we expand
Eqs. (\ref{final_homo_eq_simplified},\ref{final_homo_eq_simplified_S}) near $R=0$
under the same assumption as before, namely that the terms proportional to $e^S$ may be ignored
(this requires that $e^S<< \dot{R}^2/R$, which will again be justified a posteriori). We obtain \bea&\ddot{R}  =\frac{\dot{R}^2}{2R}, \label{IS_Eq_K=0}\\
&\ddot{S}=-\frac{\dot{R}^2}{2R^2}.\label{IS_Eq_K=0_S} \eea These
equations are easily solved: \bea & R\left( t \right) = C_1 \left|
{t - t_0 } \right|^2  \label{R=0 solution}\eea and \bea & S\left(
t \right) = C_2  + C_3 |t-t_0| + 2\ln \left| {t - t_0 },
\right|\label{R=0 solution S}\eea where $C_1, C_2, C_3, t_0$ are
four arbitrary parameters. We find that the present situation
differs from the $R=K$ singularity, because now $\dot{R}$ (like
$R$) is finite and $S$ (like $\dot{S}$) diverges. Nevertheless,
our preassumption $e^S<<\dot{R}^2/R$  is still justified, because
$e^S$ vanishes as $(t-t_0)^2$ whereas $\dot{R}^2/R$ approaches a
constant.

It is quite surprising to find that although the evolution equations are singular at $R=0$, the evolving solution for both $R$ and
$e^S$ is regular there. In particular, the metric function $\rho=(S-\ln R)/2$ is finite at $R=0$. Note, however, that the dilaton $\phi$
diverges there.

Our homogeneous solution starts at $R=R_{0}>>K$, and $R$ decreases until the $R=K$ singularity is reached.
The divergence of $\dot{R}$ and $\dot{S}$ at $R=K$ results
in an inability to provide a unique prediction for the evolution of the fields beyond that point.
Therefore the $R=0$ singularity and its neighborhood are beyond our domain of prediction.
In the rest of this paper we shall concentrate on the $R=K$ singularity,
as well as on the global behavior in the domain $K<R<M$.

\section{Constructing the interior homogeneous solution}

So far, we dealt with the local structure of the fields near the singularity.
Our ultimate goal, however, is to understand the
global behavior of the fields inside the BH. To this end, we need to
follow the evolution of $R(t)$ and $S(t)$ from the initial hypersurface (say at $R\approx M$)
up to the singularity at $R=K$.
We shall do this by designing a couple of analytical approximations, the union of
which cover the entire domain $K<R<M$.
We shall also augment the analytic approximations by direct numerical
integration of the homogeneous evolution equations
(\ref{final_homo_eq_simplified},\ref{final_homo_eq_simplified_S}).
We shall start by specifying the initial conditions, which are
required for both the numerical and analytical solutions.

\subsection{Setting the initial conditions}
The required initial conditions for the homogeneous solution are the four functions $R,\dot{R},S,\dot{S}$, all set at
a certain initial moment $t_0$. The basic strategy of setting the initial data follows from the discussion in section \ref{sec. Homogenous_equation}.
Suppose that we want to explore the interior of the evaporating BH at a stage (i.e. $v$ value)
where its remaining mass is $M$ (with $M>>K$). Then we take the initial conditions to be those corresponding to
the classical internal Schwarzschild solution with the same mass parameter $M$,
at a hypersurface $R=const\equiv R_{0}$, at the limit $R_{0}\to M$.
From Eq. (\ref{schwartzschild_sol_eq}) this amounts to setting
$R=M$, $\dot R=0$, $\dot S=1$, and $S=t_{0} \to -\infty$.
In actual numerical implementations we pick a large negative value of $t_0$ (say, $t_0=-30$) and set $S=t_0$.
Note that in such a horizon-limit setup of initial conditions the choice $\dot R=0$ is crucial for regularity.
Any other choice of initial $\dot R=0$ at $t \to -\infty$ will lead to a solution which lacks a regular horizon, and is hence inappropriate
for approximating the spacetime of an evaporating BH.
\footnote {The homogeneous evolution equations (\ref{final_homo_eq_simplified},\ref{final_homo_eq_simplified_S})
admit a generic class of solutions in which  $\dot R$ approaches
a negative constant as $t \to -\infty$. However in this class obviously  $R\rightarrow\infty$ at that limit,
and the curvature diverges as well.}

For this initial-value setup, the constraint equation (\ref{semi-classical constraint}) implies that $\hat T$ vanishes
at the horizon limit $t \to -\infty$.
Eq. (\ref{constarint}) then yields $\hat z=1/4$.
\footnote{This is the unique $\hat z=\hat z_{v}=\hat z_{u}$ value in a precisely homogeneous solution with a regular horizon.
We point out, however, that the actual evaporating-BH solution is not precisely homogenous,
and one finds that at the horizon limit $\hat z_{u}=1/4$ and $\hat z_{v}=0$.
This difference between $\hat z_{u}$ and $\hat z_{v}$ may be regarded as a measure for the deviation of the
actual evaporating-BH spacetime from the precisely homogeneous solution. Simple analytical arguments suggest that the effect of this change in $\hat z$ by $1/4$ will be small
for large $M/K$ (and vanish at the macroscopic limit  $K/M \to 0$). We explored this effect numerically
(within the homogeneous framework), for $K/M$ ranging between $0.01$ and  $0.001$, and verified
that this is indeed the case. As a quantitative measure for this effect,
one can look at the change in the actual $B$ parameter at the $R=K$ singularity,
induced by such a change of $1/4$ in $\hat z$. Our numerical results suggest that the fractional change in $B$
scales as $K/M$.}

\subsection{Approximate solutions in the different regimes}
We need to evolve the fields $R,S$ from the initial hypersurface, which we set to be at $R\rightarrow M$, up to the singularity at
$R=K$ (recall we assume $M>>K$). We notice three important domains in this overall range :
 (i) the macroscopic domain $R>>K$. (ii) very close to the $R=K$ singularity, and (iii) the domain $K<R<<M$
[Note that (iii) is an extension of (ii), and it overlaps with (i).] We shall now discuss the approximate solution
in each of these domains.

\subsubsection{The macroscopic (or classical) domain}
It is easy to see that in the domain $R>>K$ the quantum contribution to the right-hand side of
Eq. (\ref{final_homo_eq_simplified}) is negligible compared to the classical contribution ($K=0$).
The same applies to Eq. (\ref{final_homo_eq_simplified_S}).
Therefore, in this domain the semiclassical solution may be well approximated by the classical
solution (\ref{schwartzschild_sol_eq}). (For numerical verification see Fig. \ref{fig_EGS} below.)

\subsubsection{Near the singularity}
This domain is characterized by the inequality $R-K<<K$.
The approximate solution in this region was constructed
in the previous section, Eqs. (\ref{Solution_at_K_R},\ref{Solution_at_K_S}). This solution is
characterized by four parameters, of which the most significant one is $B$.

\subsubsection{The small-$R$ approximation ($R<<M$)}

We seek an intermediate approximate solution, in the domain
where $R$ is $<<M$, yet it is not quite close to $K$ (say, $R \sim 3K$).
Formulating the desired approximation, that will apply both near the
singularity and in the portion $R<<M$ of the classical domain,
requires understanding the common basis of these two approximations.
Consider the expression given in the right-hand
side of Eq. (\ref{final_homo_eq_simplified}) for $\ddot{R}$.
In the near-singularity approximation the first term in the
right-hand side (which is $\propto e^S$) is negligible compared to the second one since
$\dot{R}$ diverges, whereas $S(t)$ is bounded as seen in Eq.
(\ref{Solution_at_K_S}). One observes, however, that this term
remains unimportant even in the portion $R<<M$ of the classical
domain. In this regime, both $\dot {R}$ and $-e^{S}$ are given
approximately by $ R-M \cong -M$, see Eq. (\ref{Schwarzschild_sol_derivative}), hence $e^{S}/ \dot {R}^2$
scales as $1/M$. Therefore, if we fix $R$ and increase $M$, the
first term in the right-hand side of Eq. (\ref{final_homo_eq_simplified}) becomes negligible---just like in
the near-singularity approximation. Omitting this term we obtain: \bea \ddot{R} = - {\dot{R}}^2 \frac{K}{{2R\left( {R - K} \right)}}.
\label{LMA_equations} \eea We shall refer to this approximation as
the {\it small-$R$ approximation}.
Note that the near-singularity approximation may be obtained from the
small-$R$ approximation by substituting $R\cong K$,
as may be seen by comparing Eqs. (\ref{LMA_equations}) and (\ref{IS_equations}).

Solving Eq. \ref{LMA_equations} for $\dot{R}$ we obtain \bea \dot{R}= A\sqrt{\frac{R}{R-K}},\label{LMA_solution} \eea
where $A$ is a free parameter. One more simple integration yields an expression for $t(r)$, which we do not need however.

In a particular evaporating-BH space-time, the parameter $A$ in
Eq. (\ref {LMA_solution}) is to be determined from the effective mass parameter $M$. This can be done by matching
the small-$R$ approximation and the classical approximation in their
overlap domain $K<<R<<M$.
Eqs. (\ref{LMA_solution}) and (\ref{Schwarzschild_sol_derivative}) respectively
yield $\dot{R} \cong A$ and $\dot{R} \cong -M$ in this domain, implying \bea A \cong -M. \label{A-M}\eea

The small-$R$ approximation was established above by justifying it in
both edges of its domain of validity (namely at $R\cong K$ and at $K<<R<<M$). We still need to demonstrate
its validity in the region in between, where $R$ is of order a few times $K$.
To this end we note from Eqs. (\ref{LMA_solution},\ref{A-M}), that for fixed $R$ and $K$, $\dot{R}$ scales as $M$
(just like $e^S$). Therefore, in the right-hand side of Eq. (\ref{final_homo_eq_simplified})
the second term scales as $M^2$, whereas the first term scales as $M$ and can therefore
be neglected at the large-$M$ limit ($M>>K$).
This justifies the small-$R$ approximation a posteriori,
in its entire domain $K<R<<M$.

\begin{figure}[htb]
\begin{center}
\subfigure[]{\includegraphics[scale=0.6]{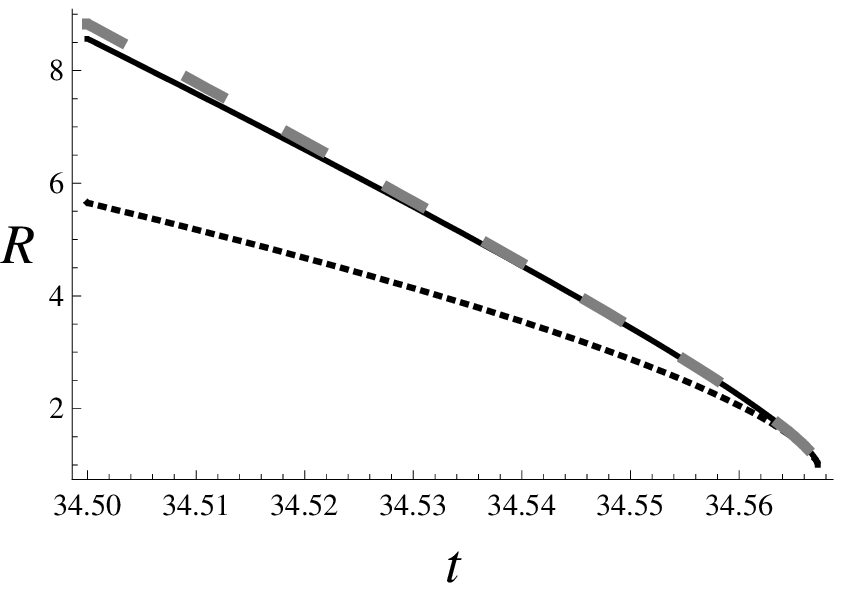}}\
\subfigure[]{\includegraphics[scale=0.6]{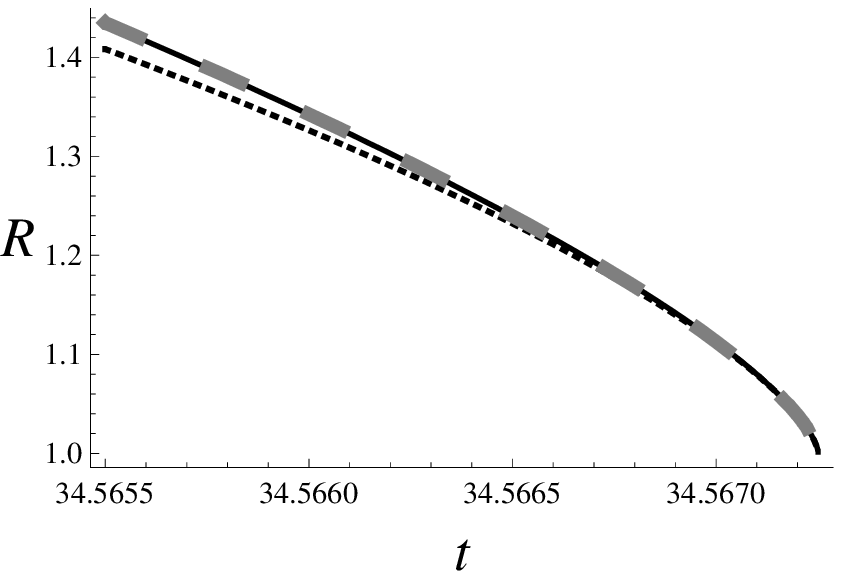}} \caption{Plots of
$R(t)$ for the full numerical solution (solid black), the
near-singularity approximation (dotted black) and the small-$R$
approximation (dashed gray). All plots are for $K=1$ and $M=10^2$.
(a) The domain where $K<R<8K$. The large deviation of the
near-singularity approximation is apparent. (b) A zoom at the
domain closer to the singularity, where $K<R<1.4K$.}
\label{fig_match}
\end{center}
\end{figure}

The validity of the small-$R$ approximation, and its effectiveness compared to the near-singularity approximation,
is demonstrated numerically in Fig. \ref{fig_match}.

Finally, we can now determine the near-singularity coefficient $B$
in Eq. (\ref{Solution_at_K_R}), as a function of the effective
mass $M$. Differentiating Eq. (\ref{Solution_at_K_R}) yields
$\dot{R}=-\frac{2}{3}B^{3/2}\sqrt{1/(R-K)}$. Substituting Eq. (\ref{A-M}) and $R\cong
K<<M$ in Eq. (\ref{LMA_solution}) yields $\dot{R}=-M
\sqrt{K/(R-K)}$. Matching the two last expressions for $\dot{R}$,
we obtain \bea B=(\frac{3}{2}\sqrt{K}M)^{\frac{2}{3}}.\label{B
relation to IC}\eea This analytic expression for $B$ (valid at the
leading order in $K/M$) may be verified numerically by evaluating
$\dot R \sqrt{R-K}$ at the limit where $R \to K$. For $K/M$ values
of $0.01$, $0.003$, and $0.001$, we numerically find the fractional
deviation of numerical $B$ from (\ref{B relation to IC}) to be  $0.016$, $0.006$, and $0.002$, respectively---suggesting
that this fractional deviation may scale as $K/M$.

\subsection{Global approximate solution}

A global approximate homogenous solution in the range $K<R<M$ can be derived by merging the
classical approximation $\dot{R}\cong R-M$ and the small-$R$ approximation
$\dot{R}\cong -M\sqrt{\frac{R}{R-K}}$  to obtain a single approximate
expression: \bea \dot{R}\cong
(R-M)\sqrt{\frac{R}{R-K}}.\label{EGS_equation} \eea
Substituting $R>>K$
in Eq. (\ref{EGS_equation}) yields the classical approximation.
Substituting $R<<M$
yields the small-$R$ approximation (\ref{LMA_solution}) (recall $A=-M$).
Since the union of the domains $R>>K$ and
$R<<M$ covers the entire range of integration $K<R<M$ (recall that
we assume $M>>K$ throughout), the approximation
(\ref{EGS_equation}) is valid everywhere.
We shall refer to it as the {\it large-mass approximation}.

\begin{figure}[htb]
\begin{center}
\subfigure[]{\includegraphics[scale=0.6]{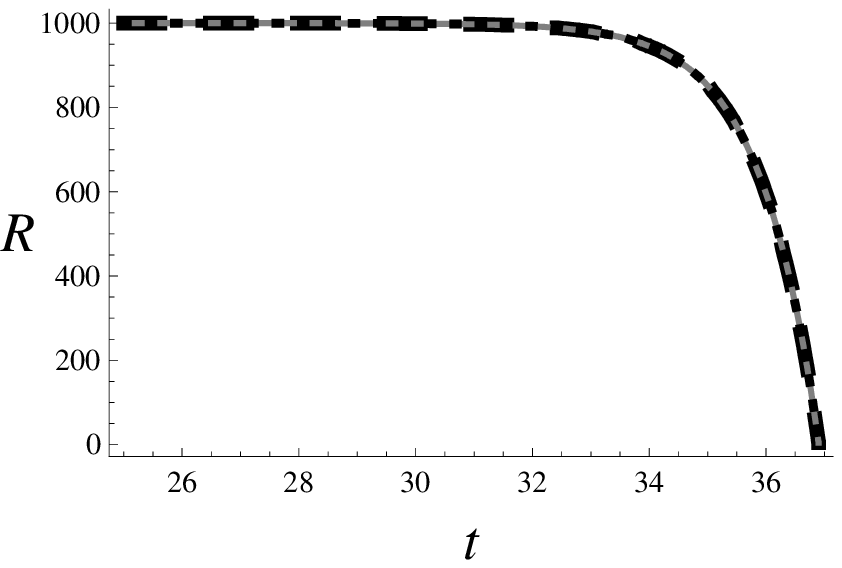}}\
\subfigure[]{\includegraphics[scale=0.6]{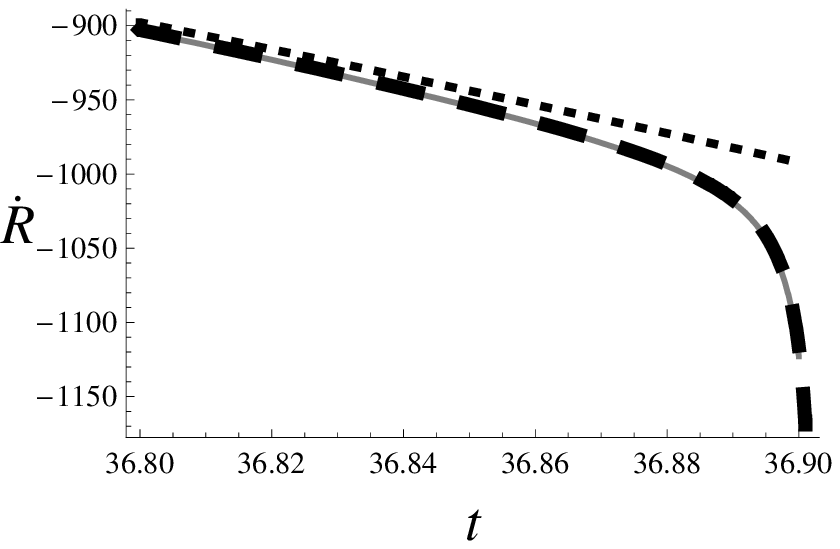}} \caption{ Plots
of $R$ (a) and $\dot{R}$ (b) for the full numerical solution
(dashed black), for the global large-mass approximation (solid
gray), and for the classical Schwarzschild solution (dotted
black), for $K=1$ and $M=10^3$. The breakdown of the classical
approximation for $\dot{R}$ at small $R$ is evident. The validity
of the large-mass approximation throughout the entire domain
$K<R<M$ is clearly demonstrated.} \label{fig_EGS}
\end{center}
\end{figure}

A comparison of the large-mass approximation
and the full numerical solution, in the entire domain $R<M$, is presented in
Fig. \ref{fig_EGS}.

\section{Effective Lagrangian and Hamiltonian for the near-singularity region}

The occurrence of a spacelike singularity at $R=K$ makes it difficult (if not impossible) to explore
the evolution of physics---and spacetime---beyond $R=K$ by semiclassical methods.
In Ref. \cite{Ashtekar} it was suggested that quantum evolution will still be regular even
at the (would-be) semiclassical singularity. This motivates one to
study the evolution of $R$ and $S$ in the neighborhood of the singularity
in a quantum-mechanical framework.
One of the simplest options would be to address the quantum problem
at the "mini-superspace" level: Since at the semiclassical level the homogeneous solution
provides a good approximation to the actual spacetime evolution (as long as $M>>K$),
one may drop all spatial derivatives and analyze the simpler problem in which $R$ and $S$
depend on $t$ solely.
Furthermore, since the critical stage of evolution is the transition across the $R=K$
singularity, it may be sufficient (at least as a first step) to analyze the problem at the leading
order in $R-K$.

This problem of quantum evolution is certainly beyond the scope of the present paper. Nevertheless, we do
take here the preparatory steps which are to be implemented already at the semiclassical level:
Namely, the formulation of an effective Lagrangian (and subsequently an effective Hamiltonian)
that describes the evolution of $R(t)$ and $S(t)$ near $R=K$. We therefore seek a Lagrangian
$L(R,S,\dot R,\dot S)$ that will recover the
evolution equations (\ref{IS_equations},\ref{Eq_IS_S}) near $R=K$.

Since Eq. (\ref{IS_equations}) constitutes a closed equation of motion for $R(t)$, we start by
constructing an effective Lagrangian $L(R,\dot R)$ for this equation.
This turns out to be
\bea L_{R}=\frac{1}{2}(R-K)\dot{R}^2,\eea
as one can easily verify by applying to it the Euler-Lagrange equation.
The momentum conjugate to $R$ is
\bea P_R=(R-K)\dot{R}. \eea
This yields the effective Hamiltonian for the $R$-field:
\bea H_{R}=\frac{1}{2}\frac{P_R^2}{R-K} \eea

\footnote{The effective Hamiltonian $H_{R}$ may also be obtained in a more direct manner
as follows: One starts from the original Lagrangian density (\ref{Action}), omits all spatial derivatives
(namely, replace $\partial_v$ and $\partial_u$ by a $t$-derivative),
omit the $f_{i}$ contribution, and transform $\phi,\rho$ to $R,S$, to obtain a
Lagrangian $L(R,S,\dot R,\dot S)$.
Then one constructs from $L$ the conjugate momenta $P_{R},P_{S}$ and the Hamiltonian
$H(R,S,P_{R},P_{S})$.  One then expands $H$ in powers of $R-K$.
The effective Hamiltonian $H_{R}$ is obtained (up to a multiplicative constant)  as the leading order in this expansion.}.

Next we treat the combined system of $R$ and $S$ (still in the homogeneous framework and at the
leading order in $R-K$). To this end we find it convenient to replace $S$ by the new variable $Z=R+KS$.
From Eqs. (\ref{IS_equations},\ref{Eq_IS_S}) it follows that $\ddot{Z}=0$.
Hence $Z(t)$ corresponds to a free one-dimensional motion, with
$L_{Z}=\frac{1}{2}\dot{Z}^2$, conjugate momentum $P_Z=\dot{Z}$, and Hamiltonian
$H_{Z}=\frac{1}{2}P_Z^2$. The overall near-singularity Lagrangian and Hamiltonian are $L=L_R+L_Z$ and $H=H_R+H_Z$.

\section{Discussion}

In this manuscript we primarily discussed two closely related
issues. The first is the structure and main properties of the
spacelike singularity which forms inside a CGHS \cite{CGHS}
evaporating BH. The second is the structure of the fields
(spacetime and dilaton) in the entire range of the BH interior,
from the horizon up to the singularity.

As a first step we established the homogenous approximation, a
central tool which allowed us to analyze the above issues by means
of ordinary differential equations. This approximation is valid as
long as the BH mass is macroscopic.

We find that the field equations admit two singular points: at $R=0$ and $R=K$. Using the
homogenous approximation, we wrote down the asymptotic form of the
field equations near each of these singularities [Eqs.
(\ref{IS_equations},\ref{Eq_IS_S}) and
(\ref{IS_Eq_K=0},\ref{IS_Eq_K=0_S})] and solved them exactly,
obtaining the asymptotic solutions
(\ref{Solution_at_K_R},\ref{Solution_at_K_S}) and (\ref{R=0
solution},\ref{R=0 solution S}) at the two singularities.
In the scenario of BH formation, $R$ starts at large values and then decreases, so the $R=K$ singularity will be reached first.
It is therefore not clear whether the $R=0$ singularity will ever form.
Hence, in the rest of our analysis we concentrate on the $R=K$ singularity.

Next we explored the evolution of the fields inside the BH from
the horizon up to the singularity. By combining several
approximations, we composed a global approximation for $\dot{R}$
as a function of $R$, given in Eq. (\ref{EGS_equation}).
This expression can be further integrated in a
straightforward manner to yield an expression for $R(t)$ (though in a
functionally-implicit form). It should also be possible to use this
expression in order to construct an approximate global solution
for $S(t)$, but this requires some technical work which is beyond the scope
of this manuscript \footnote{The specific form of $S(t)$ is less relevant here
than that of $R(t)$, since the singularity is determined by the value of $R$ but insensitive to $S$.}.

The overall approximate expression (\ref{EGS_equation}) for $\dot{R}(R)$  allows us to determine
the parameter $B$ of Eq. (\ref{Solution_at_K_R}), which was left
as a free parameter by the local analysis near $R=K$. This
parameter is found to depend on the BH's effective mass $M$,
through Eq. (\ref{B relation to IC}).

Our analysis made an extensive use of the local approximate homogeneity which characterizes the $R=K$ singularity.
One should bare in mind, however, that the overall, large-scale, structure of the singularity is inhomogeneous after all.
The parameter $B$, which characterizes the local singularity structure,
slowly drifts along the $R=K$ singularity line of Fig.\ref{fig_space_time}.
This is explicitly seen by substituting $M=M(v)$ in Eq. (\ref{B relation to IC}).
(A similar drift may also apply to the other local parameters $B_2$, $B_3$, $t_0$.)
This qualitative picture holds as long as the remaining BH mass is macroscopic, namely $M(v)>>K$.

From the asymptotic solution
(\ref{Solution_at_K_R},\ref{Solution_at_K_S}), one immediately
observes that $R$ and $S$ are continuous at $R=K$, yet their
derivatives diverge there.  The same  applies to $\rho$. Thus, the metric tensor is continuous and non-singular, but its derivative diverges. The $R=K$ singularity may be classified as \textit{deformationally-weak} \cite{Tipler_singularity},\cite{Amos_singularity}: An extended body will only experience a finite tidal deformation as it approaches the singularity.

 We briefly discuss here the possibility of extending
semiclassical spacetime beyond the $R=K$ singularity. From the
discussion above it is obvious that a differentiable extended
spacetime is ruled out, because differentiability already breaks
down on the approach to $R=K$ from the past. On the other hand,
continuous extensions do exist. In fact, there is a two-parameter
family of such continuous extensions (characterized by the values of B
and $B_{2}$ beyond $R=K$). Thus, in our view, the problem of
extending the semiclassical solution beyond the singularity is
primarily a problem of \textit{ambiguity}: A priori it is not clear which
of the infinite possible extensions will be chosen by Nature (if
any). We expect that a fully-quantized treatment (like \cite{Ashtekar}
for example) should provide the answer to the extension problem:
First, it should tell us if a semiclassical phase at all develops
beyond the $R=K$ singularity, and second, if such a semiclassical
extension indeed develops, it should determine which semiclassical
branch is selected.

Finally we constructed an effective Lagrangian and Hamiltonian
near the $R=K$ singularity. Such an effective Hamiltonian may be a
useful input for future quantum-mechanical treatments of the
behavior near the singularity.

This research was supported by the Israel Science Foundation
(grant no. 1346/07)


\end{document}